\documentclass[%
 preprint,
 amsmath,amssymb,
 aps,
]{revtex4-1}

\usepackage{rotating}
\usepackage{epsfig}
\usepackage{setspace}

\usepackage{graphicx}
\usepackage{bm}
\usepackage{epstopdf}

\begin{document}

\title{Study of Nb$_{0.18}$Re$_{0.82}$ non-centrosymmetric superconductor in the normal and superconducting states}
\author{Shyam Sundar$^{*,1}$, S. Salem-Sugui Jr.$^{1}$, M. K. Chattopadhyay$^{2,3}$, S. B. Roy$^{^\ddagger,2}$, L. S. Sharath Chandra$^{2}$, L. F. Cohen$^{4}$, L. Ghivelder$^{1}$}
\affiliation{$^1$Instituto de Fisica, Universidade Federal do Rio de Janeiro, 21941-972 Rio de Janeiro, RJ, Brazil}
\affiliation{$^2$Free Electron Laser Utilization Laboratory, Raja Ramanna Centre for Advanced Technology, Indore-452 013, India}
\affiliation{$^3$Homi Bhabha National Institute, Training School Complex, Anushaktinagar, Mumbai 400 094, India}
\affiliation{$^4$The Blackett Laboratory, Physics Department, Imperial College London, London SW7 2AZ, United Kingdom}

\begin{abstract}

We examine the evidence for multiband superconductivity and non s-wave pairing in the non-centrosymmetric superconductor Nb$_{0.18}$Re$_{0.82}$, using electrical transport, magnetization and specific heat measurements. In the normal state, both the  evolution of resistivity with temperature and with magnetic field support a multiband picture. In the superconducting state, the Werthamer, Helfand and Hohenberg (WHH) model cannot adequately describe the temperature dependence of the upper critical field, $H_{c2}(T)$, over the whole temperature range measured. In addition, the observed $H_{c2}(0)$ exceeds the Pauli limit, suggesting non-s-wave pairing. Interestingly, the Kadowaki-Woods ratio and Uemura plot reveal a behavior in Nb$_{0.18}$Re$_{0.82}$ which is similar to that found in unconventional superconductors. The temperature dependence of the lower critical field, $H_{c1}(T)$, follows an anomalous $T^3$ behavior and the derived normalized superfluid density ($\rho_s$) is well explained using a nodeless two-gap description. Phase-fluctuation analysis conducted on the reversible magnetization data, reveals a significant deviation from the mean-field conventional s-wave behavior. This trend is interpreted in terms of a non s-wave spin-triplet component in the pairing symmetry as might be anticipated in a non-centrosymmetric superconductor where anti-symmetric spin-orbit coupling plays a dominant role. 

\end{abstract}


\maketitle

\section{Introduction}

The inversion and the time-reversal symmetry of the ground state of the superconducting wave-function defines the pairing state of the Cooper pairs, which may be categorized in terms of even-parity spin-singlet or odd-parity spin-triplet pairing \cite{Ern12}. The lack of either one of these symmetries in a system gives rise to unconventional superconductivity. The lack of inversion symmetry is controlled by the strength of anti-symmetric spin-orbit coupling (ASOC), which leads to the spin splitting of the electronic states at the Fermi level and may give rise to a mixture of spin-singlet and spin-triplet pairing states in the superconducting wave function \cite{Ern12, kne15}. In addition, non-centrosymmetric superconductors (NCS) can also exhibit a variety of other  unconventional properties, such as, magneto-electrical effects, multigap behavior, helical vortex-state, nodes in the superconducting gap, non-trivial topological effects and time reversal symmetry breaking (TRSB) \cite{smi17}. These copious exotic behaviors in NCS are topical to investigate fundamentally and could even underpin the development of new technological device concepts, such as, those based on Majorana Fermions \cite{car16, Ern12,kne15,smi17}.

The discovery of superconductivity in a heavy-Fermion non-centrosymmetric compound, CePt$_3$Si, ignited the rapid growth in the field of NCS \cite{bau04, kim05, aka04, smi17}. Most of the heavy Fermion NCS show superconductivity under finite applied pressure (except CePt$_3$Si) and the effect of strong electron correlations and spin-fluctuations are responsible for the observation of unconventional superconductivity in these systems \cite{kim12, smi17}. Interestingly, due to the absence of magnetic correlations, the weakly-correlated electron systems, such as, Nb-Re \cite{kar11}, LaNiC$_2$ \cite{sus17}, Li$_2$(Pd,Pt)$_x$B \cite{tog04, pet05}, (Rh,Ir)Ga$_9$ \cite{tak07} and Mg$_{10+x}$Ir$_{19}$B$_{16−y}$ \cite{kli06}, are particularly interesting because they better enable the isolation of the impact of ASOC in NCS. Most of the weakly-correlated electron systems appear to show s-wave behavior, although, in some cases, they also show nodal superconducting gap structure \cite{smi17, kne15, lan17}. It has also been observed that even in the case of fully gaped s-wave behavior, $\mu$-SR measurements found TRSB in La$_7$Ir$_3$, LaNiC$_2$ \cite{bar15, hil09}.

Nb$_{0.18}$Re$_{0.82}$ is a phonon-mediated non-centrosymmetric superconductor having superconducting transition temperature, $T_c$, in the vicinity of 9 K \cite{kar11}. Literature suggests that electronic correlations are not strong and magnetic correlations are also absent, which makes this system  suitable to investigate the role of ASOC \cite{kar11, che13}. It is also observed that the upper critical field, $H_{c2}$, in Nb$_{0.18}$Re$_{0.82}$ superconductor reaches the Pauli limit, which suggests the possibility of unconventional pairing \cite{kar11, che13}. A recent study of the specific heat and Andreev point contact spectroscopy on a single crystal of Nb$_{0.18}$Re$_{0.82}$ concluded that the system is a nodeless two-gap superconductor \cite{cir15}. However, it  has been argued that there are a variety of different scenarios based on mixed spin singlet-triplet pairing that may also explain these observations \cite{cir15}. Hence, further investigation is required to clarify the situation.

In the present study, a detailed characterization of a polycrystalline Nb$_{0.18}$Re$_{0.82}$ superconductor is performed in the normal and superconducting state. Electrical resistivity measurements in the normal state provides evidence of phonon-assisted interband scattering. The  two-gap  picture  is  confirmed  using electrical transport and magnetization results \cite{cir15}. We find that the Kadowaki-Woods ratio and Uemura plot suggest unconventional behavior which is inconsistent with prior claims \cite{che13}. The observed upper critical field, $H_{c2}(0)$, exceeds the Pauli limit, indicating the unconventional pairing, and the lower critical field $H_{c1}(T)$ follows an anomalous $T^3$  behavior. The phase analysis we have performed using reversible magnetization data does not agree with a mean-field s-wave picture and supports a scenario where a spin-triplet component exists in the order parameter.  

\section{Experimental Details}

A polycrystalline sample of Nb$_{0.18}$Re$_{0.82}$ was prepared by melting the constituent elements (99.8\% purity of Nb and 99.99\% purity of Re, Alfa Aesar) in an arc furnace with a constant supply of 99.999\% pure Argon atmosphere. The sample was flipped and re-melted six times to ensure homogeneity. A mass loss of about 1\% was observed after melting. In addition, the as-cast sample was wrapped in Ta foil and sealed in a quartz ampoule with argon atmosphere for annealing. The annealing was performed at 800 $^{\circ}$C for 7 days, followed by slow cooling, down to the room temperature. The sample homogeneity was confirmed using energy dispersive x-ray spectroscopy (EDS). The metallographic characterization was performed using a high power optical microscope. The x-ray diffraction (XRD) of the sample was performed using a standard diffractometer (Model D2 Phaser, Bruker) using Cu-K$\alpha$($\lambda = 1.5406$ \AA) radiation in the $2\theta$ range from 20$^{\circ}$ to 80$^{\circ}$ with $\Delta2 \theta$ = 0.02$^{\circ}$.

Electrical transport, specific heat, and spin susceptibility measurements were performed at low temperatures (down to 2 K) and high magnetic fields (up to 9 T), using a Physical Properties Measurement System (PPMS, Quantum Design, USA). The temperature and magnetic field dependence of electrical resistivity, $\rho$(H,T), was measured on a rectangular parallelepiped shaped sample, using a standard four probe technique. The electrical contacts were made using thin gold wires and silver paint. The contact resistance was observed $R$ $\approx$ 0.3 $\Omega$. A widely used thermal relaxation technique was exploited to measure the temperature and magnetic field dependence of specific heat, $C$(H,T), in the superconducting as well as in the normal state. The sample was also characterized using ac-susceptibility measurements performed with 0.5 Oe excitation magnetic field and 500 Hz frequency. DC magnetization measurements  performed using a squid-vsm magnetometer (Quantum Design, USA), where, a small sample was mounted within a quartz sample holder.  

\section{Results and Discussion}

\begin{figure}
\centering
\includegraphics[height=10cm]{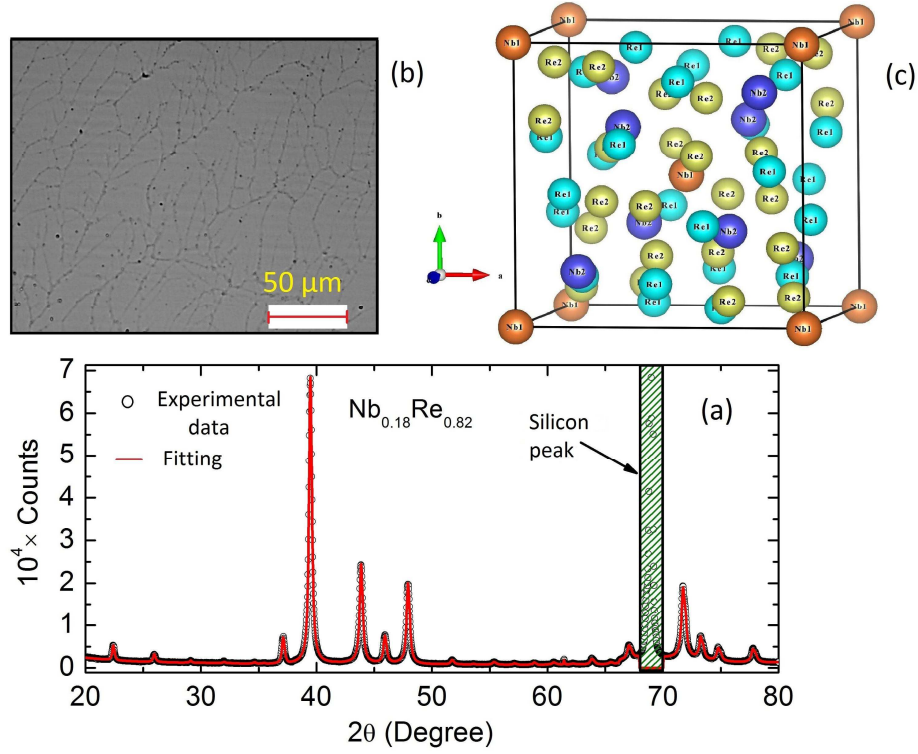}
\caption{\label{fig:xrd}(a) X-ray powder diffraction pattern of polycrystalline Nb$_{0.18}$Re$_{0.82}$, showing the $\alpha$-Mn type cubic crystal structure. The experimental data fits well to the theoretically estimated curve using Rietveld refinement method with goodness of fit, 1.4. The shaded portion shows the XRD peak related with the silicon thin film, which is used as the background in the sample holder. The silicon XRD peak is not considered as the part of the Rietveld refinement analysis. (b) The optical metallography image shows homogeneous single-phase behavior of the sample. (c) Schematic picture of the sample crystal structure (Orange spheres: 2a positions (Nb1); Blue spheres: 8c positions (Nb2); Cyan spheres: 24g1 positions (Re1); Dark yellow spheres: 24g2 positions (Re2)).}
\end{figure}

Figure \ref{fig:xrd}(a), shows the x-ray powder diffraction pattern of the polycrystalline Nb$_{0.18}$Re$_{0.82}$ sample. A silicon thin film was used as a substrate, giving a well defined XRD peak at 68.8 2$\theta$ value. The silicon XRD peak was excluded from the Rietveld refinement analysis and shown as the shaded portion in the XRD spectra. The experimental data is well explained using the $\alpha$-Mn type cubic crystal structure. The lattice parameter estimated using the Rietveld refinement analysis is 9.653 \AA, which is consistent with the literature \cite{kar11, che13}. No secondary phase peaks are observed in the XRD spectra consistent with the inspection by optical metallography shown in Fig. \ref{fig:xrd}(b), where no second  phases are observed. A schematic picture of the $\alpha$-Mn type cubic crystal structure is shows in Fig. \ref{fig:xrd}(c), where, the distribution of the Nb and Re atoms at different atomic positions are shown with different colors (See the caption of \ref{fig:xrd}(c)). 

\begin{figure}
\centering
\includegraphics[height=12cm]{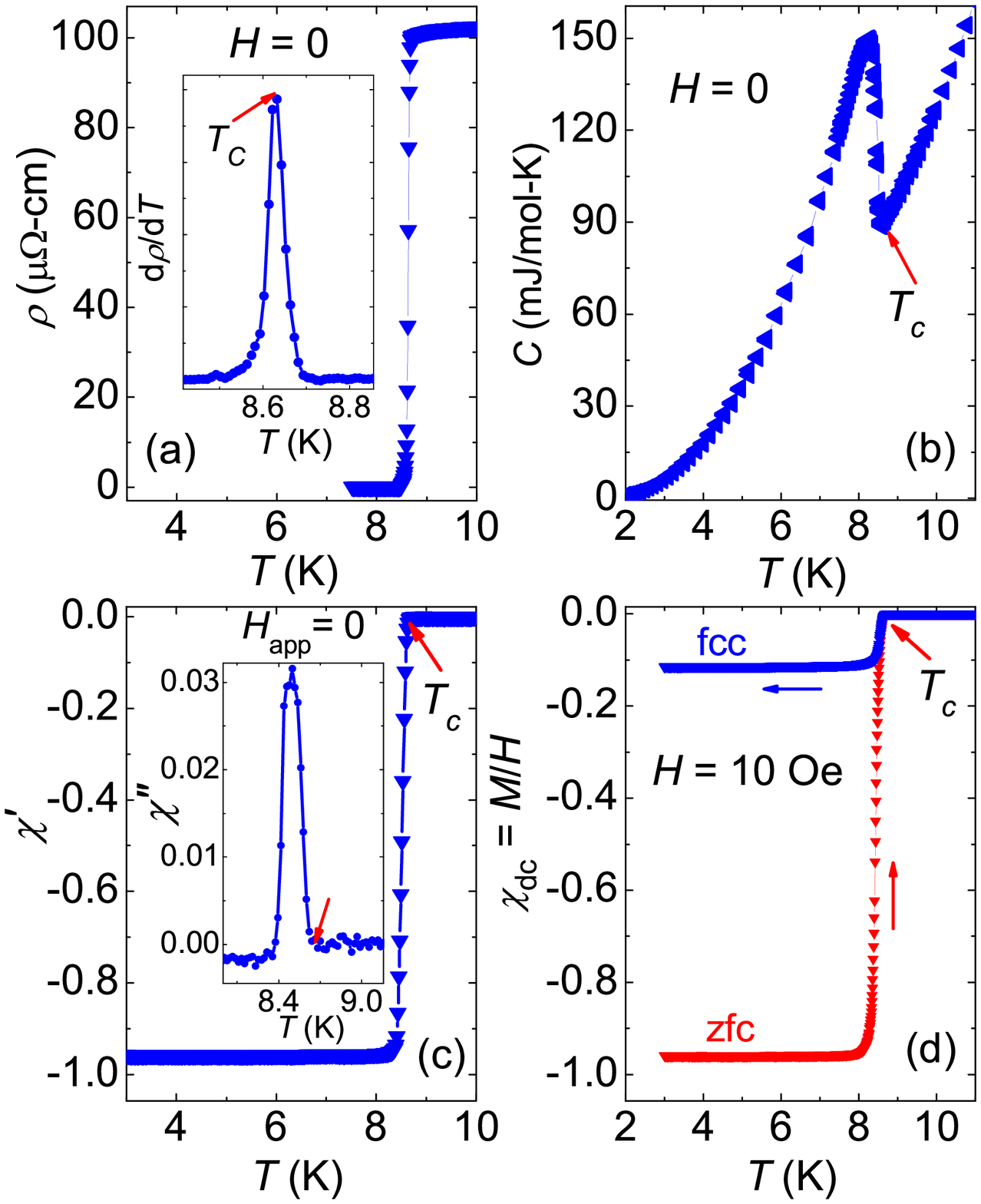}
\caption{\label{fig:Tc_all} (a) Temperature dependence of resistivity, $\rho$(T), measured in zero magnetic field, showing a sharp superconducting transition. Inset shows the derivative, d$\rho$/dT. (b) Temperature dependence of the specific heat, $C(T)$, measured at $H$ = 0. (c) Real part of temperature dependence of ac-susceptibility ($\chi'$) in zero applied magnetic field showing almost perfect shielding. The inset shows the imaginary part of ac-susceptibility ($\chi''$). (d) Temperature dependence of dc magnetization measured at $H$ = 10 Oe in zero field cooled (zfc) and field cooled cooling (fcc) protocols. The zfc curve shows the nearly perfect shielding state and the arrow mentioned the onset of $T_c$. The observed $T_c$ in all the measurements is of the order of 8.63 K $\pm$ 0.05 K.}
\end{figure}

The superconducting transition temperature is measured using various experimental techniques. In Fig. \ref{fig:Tc_all}(a), the temperature dependence of electrical resistivity, $\rho(T)$, at zero applied magnetic field shows a sharp superconducting transition with $\Delta T_c$ = 0.06 K. The inset of Fig. \ref{fig:Tc_all}(a), shows the derivative of $\rho(T)$ and the peak position is taken as the $T_c$ of the sample. Figure \ref{fig:Tc_all}(b), shows the bulk superconducting transition using temperature dependence of specific heat, $C$($T$), measured in the absence of applied magnetic field. The $\Delta T_c$ = 0.27 K, is estimated as the temperature difference between the transition onset and completion. The arrow indicates the $T_c$ of the sample. The real part of ac-susceptibility ($\chi^{'}$) in Fig. \ref{fig:Tc_all}(c), shows a sharp superconducting transition with $\Delta T_c$ = 0.15 K, where the arrow indicate the $T_c$ of the sample. The nearly perfect shielding confirms the good quality of the sample. The imaginary part of the ac-susceptibility, $\chi^"$, (inset of Fig. \ref{fig:Tc_all}(c)) shows a sharp peak. The dc-susceptibility in zfc and fcc protocols is measured at $H$ = 10 Oe, is shown in the panel (d) of Fig. \ref{fig:Tc_all}. The nearly perfect shielding and $\sim$ 12\% of Meissner fraction is indicated by zfc and fcc curves respectively and the arrow indicates the $T_c$ of the sample. All four measurements shown in Fig.\ref{fig:Tc_all} are consistent with each other and show $T_c$ = 8.63 K $\pm$ 0.05 K. The $T_c$ of the sample observed in our measurements is in agreement with the literature within the error of 0.1 K \cite{kar11, che13, cir15}. 

\begin{figure}
\centering
\includegraphics[height=12cm]{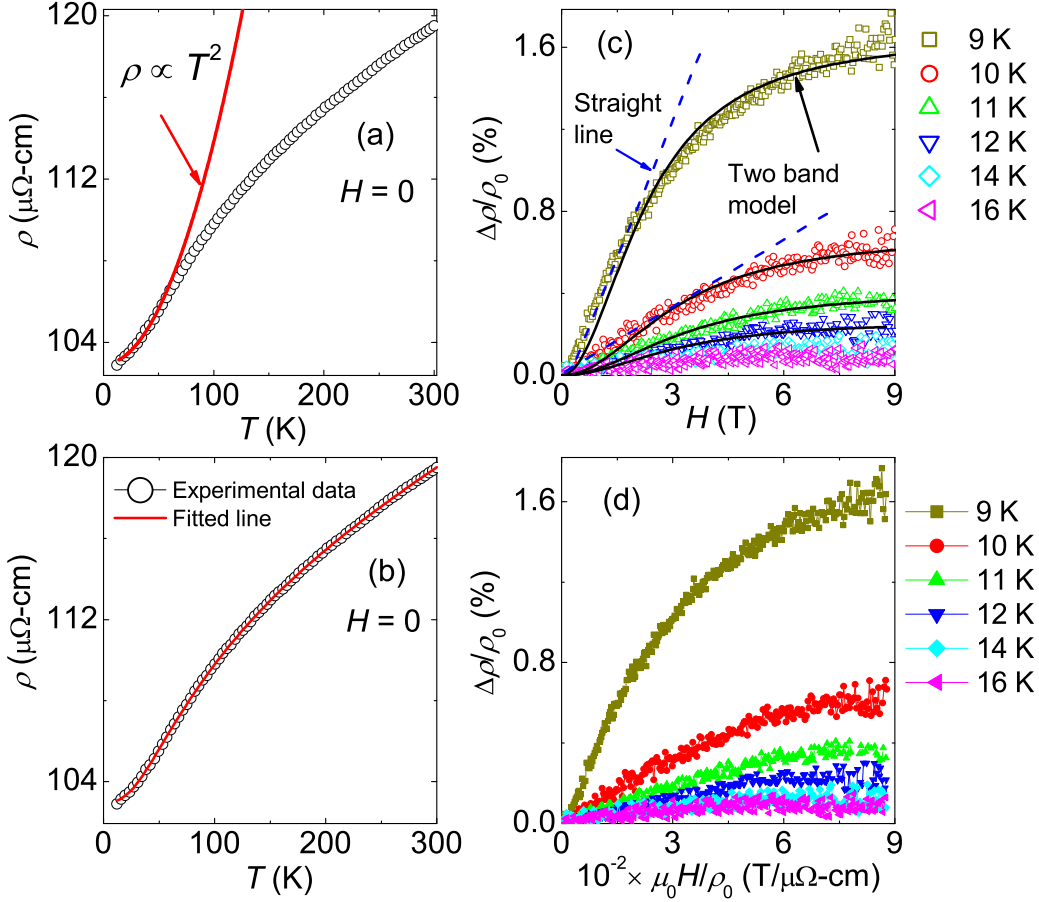}
\caption{\label{fig:RTN1} (a) $\rho(T)$ above $T_c$, showing a $T^2$ behavior in the temperature range 14-50 K. (b) An empirical relation, $\rho(T) = \rho_0 + \rho_1 T^n + \rho_2 exp(-T_0/T)$, fits the experimental data in the temperature range 11-300 K. (c) Magnetic field dependence of the transverse magnetoresistance above $T_c$, in the temperature range 9 K to 16 K. A simple two band model (solid black line) does not explain the low field experimental data, whereas, the fitting is relatively better at high magnetic fields. The dashed line shows that the experimental data varies linearly at low fields. (d)  The magnetoresistance of Nb$_{0.18}$Re$_{0.82}$ does not support the Kohler's rule.}
\end{figure}

Figure \ref{fig:RTN1} shows the electrical resistivity in the normal state as a function of temperature and magnetic field. The residual resistivity, $\rho_0$ is observed to be 102.9 $\mu$$\Omega$-cm and the residual resistivity ratio (RRR) is defined as, $\rho_{300 K}$/$\rho_{10 K}$ = 1.16. A small value of RRR suggests a dominant role of disorder in the sample. The mean free path ($l$) estimated using the Drude's theory is 2.12 \AA ~and the Pippard's coherence length obtained using the expression, $\xi_0 = 1.781\hbar \upsilon_F/(\pi^2k_BT_c)$ $\approx$ 4364 \AA, implying that the sample is in the dirty limit ($l \ll \xi_0$). Fig. \ref{fig:RTN1}(a), shows that the zero field $\rho(T)$ at low temperature (14 - 50 K) follows a quadratic behavior, $\rho = \rho_0 + AT^2$. The solid line is a fit of this expression to the experimental data, which yields the coefficient of the $T^2$ term, $A$ = 0.00109 $\pm$ 4.4E-6 $\mu$$\Omega$-cm/K$^2$. The small value of the coefficient $A$ indicates relatively weak electronic correlation in the Nb$_{0.18}$Re$_{0.82}$ alloy \cite{kar11, che13}. Figure \ref{fig:RTN1}(b) shows the zero field $\rho(T)$ where the room temperature value is 120 $\mu$$\Omega$-cm and the trend is a metal-like behavior down to 10 K. The qualitative behavior of $\rho(T)$ in the whole temperature range is similar to the one observed in Ref. [\cite{kar11}]. The experimental data is well explained using an empirical relation provided by Woodard and Cody \cite{woo64}, $\rho(T) = \rho_0 + \rho_1 T^n + \rho_2 exp(-T_0/T)$, where $\rho_0$ is the residual resistivity, $T_0$ in the exponential term defined as a characteristic temperature of a certain phonon mode. The exponential term has been explained in terms of phonon-assisted inter-band scattering or intraband umklapp scattering but the origin of non-exponential term is not defined \cite{mil76}. The empirical relation discussed above describes the data reasonably well over the whole temperature range and the estimated fitted parameters are $\rho_0$ = 102.8 $\mu$$\Omega$-cm, $\rho_1$ = 0.023 $\mu$$\Omega$-cm/K$^n$, $n$ = 1, $\rho_2$ = 12.6 $\mu$$\Omega$-cm and $T_0$ = 103.4 K. Nb$_{0.18}$Re$_{0.82}$ is considered to be a multiband superconductor, and the exponential term in the resistivity which can be interpreted in terms of phonon assisted interband scattering in the normal state, is certainly consistent with this picture. Similar  behavior has been observed in the Mo-Re multiband superconductor \cite{ss15a, ss15b}.

Figure \ref{fig:RTN1}(c), shows the magnetic field dependence of the transverse magnetoresistance, MR = ($\rho(H)-\rho_0$) $\times$ 100/$\rho_0$, measured above $T_c$, at different temperatures ranging from 9 K to 16 K. At low magnetic fields the MR follows a linear behavior and at higher magnetic fields it saturates. It is also observed that the MR does not follow the conventional metallic quadratic magnetic field dependence (MR $\propto$ $H^2$) \cite{zim72}. As Nb$_{0.18}$Re$_{0.82}$ is considered to be a multiband superconductor \cite{cir15}, we explore whether the magnetic field dependence follows the following expression for the MR in a simple two-band model \cite{zim72}.

\begin{equation} \label{eq:1}
\frac{\Delta \rho}{\rho_0} = \frac{\mu_0 H^2}{\alpha + \beta \times (\mu_0 H)^2}
\end{equation}
where, $\alpha$ and $\beta$ are the fitting parameters which are related to the conductance and mobilities of charge carriers in the associated two bands. In Fig. \ref{fig:RTN1}(c), the solid lines represent fit of Eq. (\ref{eq:1}) to the experimental data showing a good fit to the experimental data at high magnetic fields. The dotted line in Fig. \ref{fig:RTN1}(c) shows the linear dependence of the MR at low magnetic fields. The deviation between the two-band model and the experimental data may imply the contribution of more than two bands in the transport mechanism. Moreover, as the linear behavior of the MR at low fields is unexpected, we explore this further. Semi-classical transport theory based on the Boltzmann equation suggests Kohler's rule to hold in a system, if there is only one kind of charge carrier with isotropic relaxation rate in a single band system \cite{zim72}. Mathematically, the Kohler's rule may be defined as the following.
 
\begin{equation} \label{eq:2}
\frac{\Delta \rho}{\rho_0} = f(\frac{\mu_0 H}{\rho_0})
\end{equation}
where, the function $f(x)$ is temperature independent and $\rho_0$ is the residual resistivity. Equation (\ref{eq:2}) suggests that the plot of $\Delta \rho$/$\rho_0$ vs. $\mu_0 H/\rho_0$ should produce a merged universal curve for all isothermal MR measurements. In the present case, the Kohler's plot for the measured transverse MR is shown in Fig. \ref{fig:RTN1}(d). It is clearly observed that the relation is violated in all temperature and field range of measurements and indicates that the MR strongly varies with temperature. The violation of the Kohler's rule is consistent with the multiband behavior of the sample, which is seen in a previous report \cite{cir15}. Violation of the Kohler's rule is also observed in other multiband systems, such as, MgB$_2$ \cite{li06} and Mg$_{12-\delta}$Ir$_{19}$B$_{16}$ \cite{gan10}. 

\begin{figure}
\centering
\includegraphics[height=12cm]{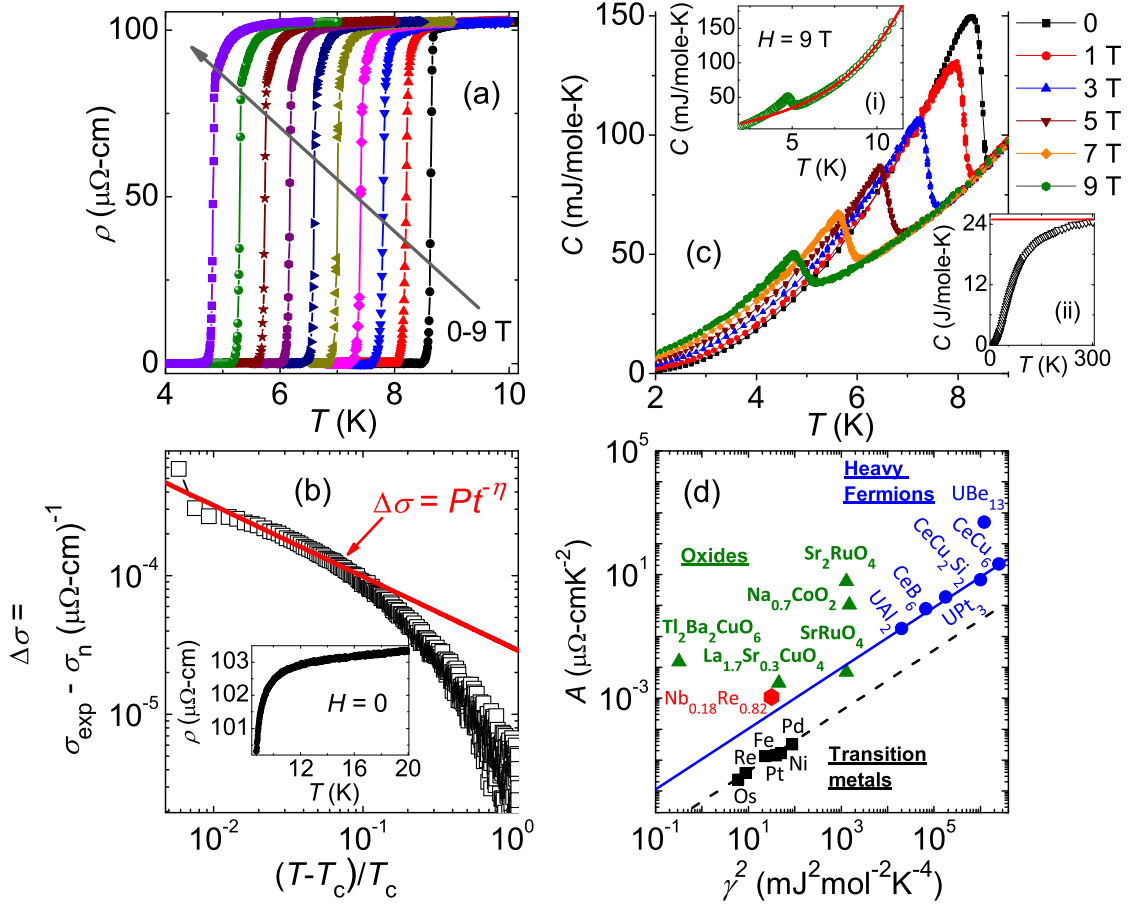}
\caption{\label{fig:RTN2} (a) Temperature dependence of the resistivity, $\rho(T)$, at different applied magnetic fields in the superconducting state. (b) The fluctuation conductivity ($\Delta$$\sigma$) as a function of ($T$-$T_c$/$T_c$) in the temperature range 8.7 K to 20 K, shows the fitting of the experimental data using the function ($\Delta \sigma_{AL} =Pt^{-\eta}$). The parameter $\eta$ $\sim$ 0.5, suggests the existence of 3-dimensional fluctuations in the sample. Inset, shows the rounding-off behavior of $\rho$($T$) at $H$ = 0, from 8.7 K to 20 K. (c) The temperature dependence of specific heat, C(T), at different applied magnetic fields. Inset (i) shows the $C(T)$ measured at $H$ = 9 T and fitted using the expression, $C(T) = \gamma T + \beta T^3 + \delta T^5$, where $\gamma$ represents the electronic part and $\beta$, $\delta$ are the phonon terms. Inset (ii) shows the specific heat measured in the normal state at zero magnetic field which approaches the Dulong-Petit value near 300 K. (d) The coefficient of the $T^2$ term $A$ in $\rho(T)$ plotted as a function of the square of the coefficient of the electronic specific heat ($\gamma^2$) for comparing different classes of materials. This plot is known as the Kadowaki-Woods (KW) plot.}
\end{figure}

Fig. \ref{fig:RTN2} (a) shows the temperature dependence of the electrical resistivity, $\rho$(T), in the superconducting state, measured in the presence of different applied magnetic fields. It is observed that the magnetic field does not play a significant role in the transition broadening. A sharp transition occurs even at high magnetic fields (9 T), with $\Delta T_c$ = 0.06 K. The superconducting transition shows a rounding behavior for zero field measurement, as well as for higher magnetic field data. This behavior at $H$ = 0, is due to the effect of thermal fluctuations \cite{mat14, ann15, han98}. Fluctuation conductivity is considered to be responsible for the observed excess conductivity in some low $T_c$ superconductors which may be understood in terms of preformed Cooper pairs well above $T_c$ \cite{mat14}. Fluctuation conductivity analysis can provide a measure of the dimensionality of the fluctuations and the coherence length in a superconductor \cite{pur93}. Fig. \ref{fig:RTN2} (b) shows the fluctuation conductivity ($\Delta\sigma = \sigma_{exp}- \sigma_n$) as a function of $(T-T_c)/T_c$ in the temperature range from 8.7 K to 20 K. The normal state conductivity ($\sigma_n$) was estimated by extrapolating the conductivity from above 3$T_c$. The inset of Fig. \ref{fig:RTN2} (b), clearly shows the rounding-off behavior for the zero field $\rho(T)$ data below 20 K. The observed experimental fluctuation conductivity ($\Delta \sigma$) can be explained in terms of Aslamazov-Larkin model, $\Delta \sigma_{AL} = Pt^{-\eta}$, where $t = (T-T_c)/T_c$, $P$ is a constant and $\eta = 2-D/2$ is a critical exponent, where, $D$ defines the dimensionality of superconducting fluctuations. In Fig. \ref{fig:RTN2} (b), the straight line fits to the experimental data ($\Delta$$\sigma$) in the temperature range 8.7 K to 9.5 K, gives $\eta$ $\approx$ 0.5, which suggests the 3D character of the superconducting fluctuations. However,the coherence length estimated using the parameter $P = e^2/32 \hbar \xi(0)$ is two orders of magnitude smaller than the reported value of $\xi(0)$ \cite{kar11}. This significant mismatch suggests that some additional mechanism plays a role in the rounding of superconducting transition near $T_c$ or perhaps it is merely due to the disorder in the sample. However, disorder usually also broaden the superconducting transition, which is not observed in the present study (see Fig. \ref{fig:Tc_all}).

The temperature dependence of the specific heat, $C(T)$, at different magnetic fields is shown in Fig. \ref{fig:RTN2} (c). In the inset (i), showing $C(T)$ measured in $H$ = 9 T, the solid line is a fit to the expression, $C(T) = \gamma T + \beta T^3 + \delta T^5$, where, $\gamma$ is the coefficient of electronic specific heat and $\beta$, $\delta$ are the phonon terms. The fitting parameters are $\gamma$ = 5.60 $\pm$ 0.04 mJ/mole-K$^2$, $\beta$ = 0.04505 $\pm$ 0.00124 mJ/mole-K$^4$ and $\delta$ = 2.50 $\times$ 10$^{-4}$ $\pm$ 8.52 $\times$ 10$^{-6}$ mJ/mole-K$^6$, which are consistent with the values obtained in the literature \cite{che13, cir15}. The $\gamma$ value shows that the electronic correlations in the sample are not strong. On the other hand, the parameter $\beta$ is used to obtain the characteristic Debye temperature ($\theta_D$), $\beta = N(12/5)\pi^4 R \theta_D^{-3}$, where, N is the number of atoms in a unit cell, R is the gas constant. For N = 1, the $\theta_D$ = 350 K and it is observed that the specific heat approaches the classical Dulong-Petit value near room temperature ($T$ = 300 K), which suggests that the higher vibrational energy modes are populated near room temperature. The specific heat jump, $\Delta C/\gamma T_c$ = 1.64 is larger than the weak coupling BCS superconductors \cite{cha07}. The electron-phonon coupling constant ($\lambda_{ep}$) is estimated using the McMillan's expression, $\lambda_{ep} = \frac{1.04 + \mu^* ln (\theta_D /1.45 T_c)}{(1-0.62 \mu^*)ln(\theta_D /1.45 T_c)-1.04}$. In this expression, $\mu^*$ is the Coulomb pseudopotential, which takes into account the direct Coulomb repulsion between electrons. For transition metals, $\mu^*$ = 0.13. Taking, $\theta_D$ = 350 K and $T_c$ = 8.63 K, the $\lambda_{ep}$ is estimated as 0.73, which is consistent with the intermediate coupling limit seen in Ref. \cite{che13} for Nb-Re superconductors. Hence, the mass enhancement is ${m_e}^* = m_e (1 + \lambda_{ep})$ = 1.73 $m_e$. The zero field $C(T)$ below $T_c$ is better explained in terms of a two-band model (see supplementary information), consistent with the single crystal study \cite{cir15}. A power law behavior in $C(T)$ below $T_c$ would indicate a non s-wave behavior. However, we do not see power law behavior in our data (see Supplementary information), nevertheless to draw a conclusion about the symmetry of the gap from this measurement alone is rather tenuous and possibly other techniques, such as, magnetic resonance probes might be needed to determine the spin symmetry of the superconducting pairing unambiguously \cite{cir15}. The quadratic dependence of $\rho(T)$ at low temperatures indicates the dominance of electronic correlations over electron-phonon scattering. In Fig. \ref{fig:RTN2}(d), the coefficient of the $T^2$ term, $A$, in zero field $\rho (T)$ is plotted against the square of the coefficient of electronic specific heat ($\gamma^2$) for Nb$_{0.18}$Re$_{0.82}$ and compared with various other classes of materials. This is widely known as the Kadowaki-Woods (KW) plot and provides a unified picture of electronic correlations in strongly correlated systems \cite{jac09}. It is clearly seen in Fig. \ref{fig:RTN2}(c), that Nb$_{0.18}$Re$_{0.82}$ follows a similar trend as heavy Fermion compounds in the KW plot and suggests that a similar emergent physics mechanism is responsible for the effective mass enhancement. This observation indicates the possibility of spin fluctuations in Nb$_{0.18}$Re$_{0.82}$ superconductor.  

\begin{figure}
\centering
\includegraphics[height=12cm]{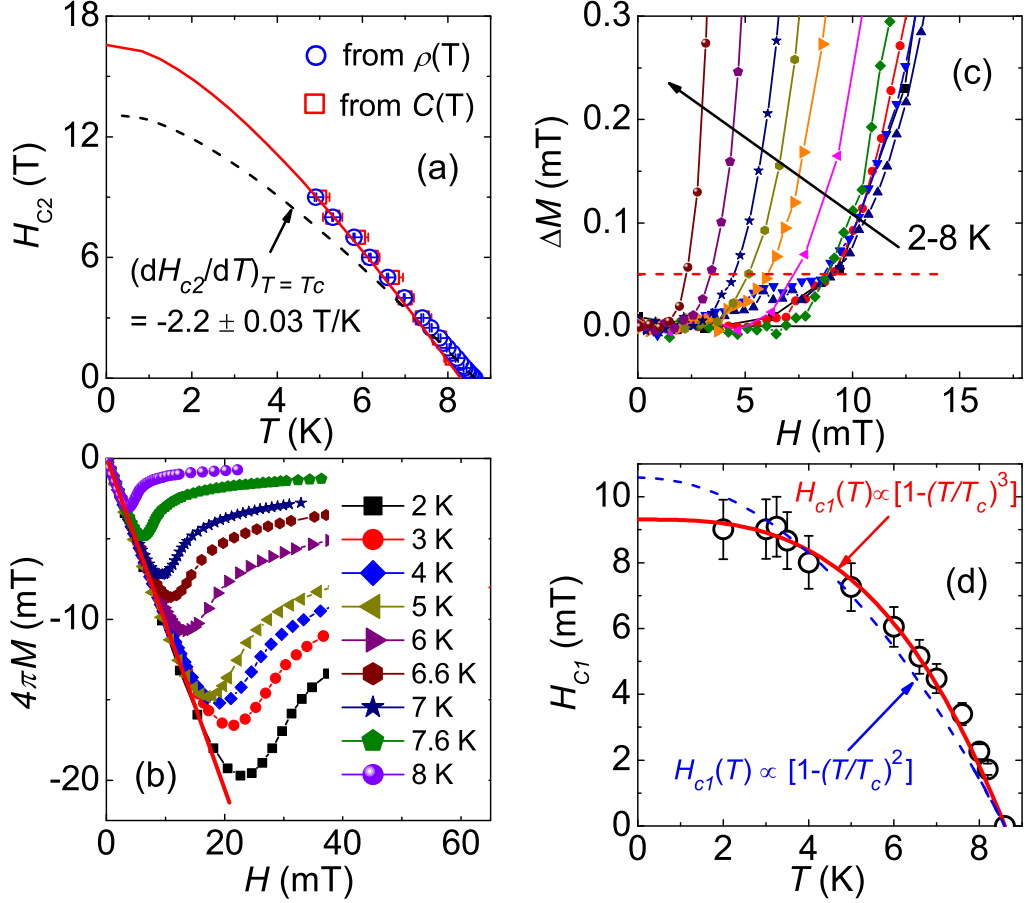}
\caption{\label{fig:HC1&2} (a) Temperature dependence of the upper critical field, $H_{c2}(T)$ estimated using temperature dependence of specific heat and resistivity measurements. The dash and solid lines show the WHH model fits well to the data in the higher and lower temperature limits respectively. (b) Isothermal $M(H)$ at low magnetic field for different temperatures below $T_c$. The $H$ values are corrected for the demagnetization effects and the straight line (slope $\approx$ -1) is fit to the experimental data to estimate $H_{c1}$. (c) The magnetic field dependence of $\Delta$$M$ in the temperature range 2-8 K, where, the $\Delta$$M$ is the difference between the experimental data $M(H)$ and the fitted straight line at low fields. The dotted horizontal line, $\Delta$$M$ = 0.05 mT, represents the criterion to estimate the lower critical field value ($H_{c1}$). (d) The temperature dependence of lower critical field, $H_{c1}(T)$. Data shows the close proximity with the $T^3$ dependence and does not follow the usual quadratic behavior.}
\end{figure}

The temperature dependence of the upper critical field, $H_{c2}(T)$, estimated from $\rho (T)$ and $C(T)$ measurements, is shown in Fig. \ref{fig:HC1&2}(a). The $T_c(H)$ is taken from the peak value in d$\rho$/d$T$ from the $\rho (T)$ data. For the $C(T)$ measurement the onset of jump is considered as the $T_c(H)$ (see Fig. \ref{fig:Tc_all}). The $H_{c2}(T)$ estimated from both $\rho (T)$ and $C(T)$ measurements are consistent with each other as may be seen in Fig. \ref{fig:HC1&2}(a). The $H_{c2}(T)$ data does not follow the empirical quadratic temperature dependence. The derivative, (d$H_{c2}$/d$T$)$_{T=T_c}$ = -2.2 $\pm$ 0.03 T/K, is estimated by fitting a straight line to few data points just below the $T_c$. The $H_{c2}(T)$ data is examined in terms of the Werthamer, Helfand and Hohenberg (WHH) model in dirty limit \cite{whh66}, which can be expressed as following. 

\begin{equation} \label{eq:3}
\ln\frac{1}{t} = \sum_{\nu = -\infty}^{\infty} {\left[\frac{1}{\left|2\nu + 1\right|} - \left\{\left|2\nu + 1 \right| + \frac{\hbar}{t} + \frac{\left(\alpha_M\hbar/t \right)^2}{\left|2\nu + 1 \right| + \left(\hbar+\lambda_{so}\right)/t}\right\}^{-1} \right]},
\end{equation}
where, $t = T/T_C$, $\hbar = 2eH(v_f^2 \tau/{6\pi T_C}) = (4/{\pi^2})H_{C2}T_C /(-dH_{C2}/dT)_{T = T_C}$. $v_f =$ Fermi velocity, and $\tau =$ relaxation time of electrons, $\alpha_M = 3/{2mv_f^2\tau} = H_{C2}(0)/{1.84 \sqrt{2}T_C}$ and $\lambda_{so} = 1/{3\pi T_C \tau_2}$, with the relaxation time of electrons for the spin-orbit interaction $\tau_2$. However, the WHH model does not adequately describe the experimental data over the whole temperature range below $T_c$. Figure \ref{fig:HC1&2}(a) shows that the WHH model (dashed line) fits the experimental data near $T_c$ but predicts an $H_{c2}(0)$ value far below the experimentally observed one. To obtain the zero field limit of the upper critical field, the WHH model is also used to fit the low temperature part of the $H_{c2}$ data giving $H_{c2}(0)$ = 16.5 T and a coherence length, $\xi(0)$ = 2314 \AA, using the expression, $\xi = (\phi_0/2\pi H_{c2})^{1/2}$. According to the WHH approach, the orbital limit of $H_{c2}(0)$ in the clean limit of a two-band superconductor and in dirty limit of a single band superconductor is 1.6 T and 1.5 T respectively. The Chandrasekhar-Clogston or the paramagnetic limit for the upper critical field, $H_p(0) = 1.84 T_c$, is estimated $\sim$ 16 T, which is smaller than the experimentally observed value of $H_{c2}(0)$. A value of the upper critical field, comparable to or larger than the Pauli limit is considered a potential signature for unconventional superconductivity \cite{clog62, cha62}. In real materials, the upper critical field of a system is generally influenced by both orbital and paramagnetic effects. The relative importance of these two competing effects may be defined in terms of Maki Parameter, $\alpha_M = \frac{H^{orb}_{c2}(0)}{\sqrt{2} H_P(0)}$, and usually $\alpha_M$ $\ll$ 1 \cite{whh66}. In case of heavy Fermions and systems with multiple small Fermi pockets, the $E_F$ may be quite small which results the $\alpha_M$ $\geq$ 1 and yields a possibility for Fulde-Ferrell-Larkin-Ovchinnikov (FFLO) state \cite{hou06, whh66}. However, in our study, $\alpha_M$ $\sim$ 0.1, which indicates the standard behavior.

We used the isothermal dc-magnetization measured at different temperatures below $T_c$ to estimate the lower critical field, $H_{c1}$. The experimental $M(H)$ data in the Meissner state is corrected for the demagnetization effect using the procedure of a linear fit near $H_{app}$ = 0. This provides the slope (= $M$/$H_{app}$) of the raw $M(H)$ data in the Meissner region. It is known that the magnetic flux density, $B$, inside a magnetic material is defined as $B = \mu_0 (H_{eff}-\alpha M)$, where, $\alpha$ is the demagnetization factor and $H_{eff} = H_{app} - \alpha M$. Hence, for a superconductor in the Meissner state, $B$ = 0, then, $M/H_{app} = 1/(\alpha - 1)$, where, $M/H_{app}$ is the slope of the raw $M(H)$ data in the Meissner state, therefore, we estimated, $\alpha$ $\sim$ 0.34. Then, the effective magnetic field is estimated as $H_{eff} = H_{app} - \alpha M$ and is plotted with the magnetization, $M$, which is shown in Fig. \ref{fig:HC1&2}(b). To estimate the lower critical field, the $M(H)$ data in Fig. \ref{fig:HC1&2}(b) is fitted with a straight line in the low magnetic field region. The deviation from linearity is considered to be the lower critical field value. The straight line fit to the low magnetic field section of the $M(H)$ curve is subtracted from the data and the resulting $\Delta M$ is plotted for different isotherms in Fig. \ref{fig:HC1&2}(c). The dashed line in the figure indicates the threshold, where $\Delta M$ decreases to the noise level, $\Delta M$ = 0.05 mT, is used as the criterion to estimate the $H_{c1}$ value. This method used widely in literature can provide a good indication of the trend of the $H_{c1}(T)$ curve although should not be relied upon for accurate absolute values \cite{con08}. Fig. \ref{fig:HC1&2}(d), shows that the temperature dependence of the extracted lower critical field, $H_{c1}(T)$ using this method, follows a cubic temperature dependence in the whole temperature range of measurement instead of the usual quadratic behavior seen previously \cite{kar11}. However, it is to be noted that in ref. \cite{kar11} the magnetization measurements were performed using vibrating sample magnetometer (VSM), whereas, in the present study, the measurements were performed using SQUID-VSM magnetometer, which has up to two order of magnitude higher sensitivity than the standard VSM. The zero field limit of the lower critical field, $H_{c1}(0)$ = 9.3 mT, is almost twice than the value observed previously \cite{kar11}. Using the expression, $\lambda (0)$ = ($\phi_0$ ln$\kappa$/4$\pi$$H_{c1}$)$^{1/2}$, the zero field limit penetration depth is estimated to be $\sim$ 305 nm, where, $\kappa$ = 192, is the Ginzburg-Landau parameter and is defined as $\kappa = H_{c2}/\sqrt{2} H_c$. The thermodynamic critical field, $H_c(0)$ = 66.4 mT, is estimated using the expression, $H_c(0) = 4.23 {\gamma}^{1/2}$$T_c$, provided in Ref.\cite{orl79}.  
   
\begin{figure}
\centering
\includegraphics[height=12cm]{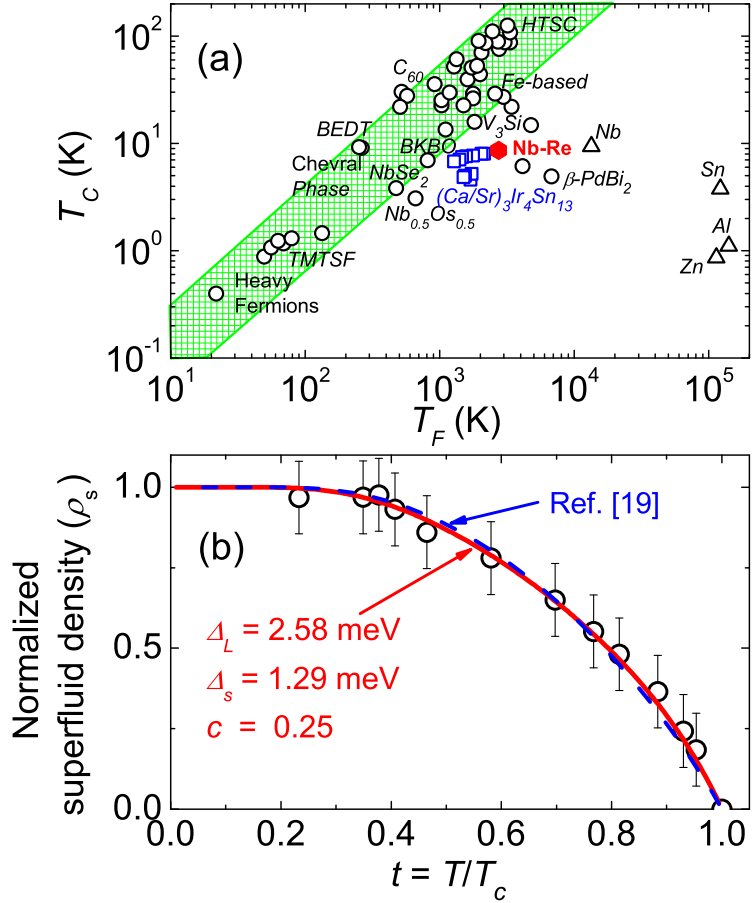}
\caption{\label{fig:SFD} (a) Uemura Plot: The superconducting transition temperature, $T_c$, plotted as a function of the Fermi temperature, $T_F$, estimated from superfluid density for Nb$_{0.18}$Re$_{0.82}$ superconductor, together with the data of other classes of superconductors (adapted from Ref. \cite{sin18,bis15}). The unconventional superconductors fall within the shaded region and the conventional superconductors lie on the right hand side of the shaded region. (b) The temperature dependence of the normalized superfluid density, $\rho_s$. The open symbols with error bars are the experimental data points and the solid line is the fit to the data by considering the two superconducting energy gaps. The dashed line is generated by using the parameters obtained in Ref. \cite{cir15} through specific heat measurement in the superconducting state.}
\end{figure}

In order to classify the non-centrosymmetric Nb$_{0.18}$Re$_{0.82}$ compound as a conventional or unconventional superconductor, we compared it with the other classes of superconductors, using the Uemura plot \cite{Uem89}, as shown in Fig. \ref{fig:SFD}(a). In the Uemura plot, the superconducting transition temperature ($T_c$) is plotted as a function of the Fermi temperature ($T_F$) estimated using the superfluid density. The shaded portion in Fig. \ref{fig:SFD}(a) represents the unconventional superconductors, such as, heavy Fermion superconductors, Fe-pnictide and high temperature superconductors. Most of the elemental superconductors, e.g Sn, Al, are well outside the shaded region. The Fermi temperature ($T_F$) for Nb$_{0.18}$Re$_{0.82}$ is obtained using the relation, $k_B T_F = \frac{\hbar^2}{2}(3\pi^2)^{3/2} \frac{n_s^2/3}{m_e(1+\lambda_{ep})}$, here, $k_B$ is the Boltzmann constant, $n_s$ is the superfluid density, $m_e$ is the electron mass and $\lambda_{ep}$ is the electron-phonon coupling constant \cite{sin18}. The superfluid density $\sim$ 5.27 $\times$ 10$^{26}$ m$^{-3}$,  is estimated using the relation, $n_s = \frac{m_e (1+\lambda_{ep})}{\mu_0 e^2 \lambda^2}$ \cite{sin18}, where, $\lambda$ is the penetration depth. The same order of magnitude is also obtained for superfluid density, using, $n_s = n_e \frac{l}{\xi_0}$, where, $l$ is the mean free path and $\xi_0$ is the Pippard coherence length. The estimated value of $T_F$ is $\sim$ 2740 K, which is plotted with the superconducting transition temperature of the sample ($T_c$) in Fig.  \ref{fig:SFD}(a). It is seen that the the data point for Nb$_{0.18}$Re$_{0.82}$ in the Uemura plot, lies in the close proximity of unconventional superconductor. A similar behavior is observed in the weakly correlated (Ca/Sr)$_3$Ir$_4$Sn$_{13}$ system \cite{bis15}. It is argued that in the presence of competing orders or multiband behavior, a phonon-mediated BCS superconductor may also shows the characteristics of an unconventional superconductor \cite{bis15}. 

In the framework of local London model, making use of the lower critical field, $H_{c1}(T)$, the normalized superfluid density, $\rho_s (T)$  is estimated using the following expression \cite{ss15a}.

\begin{equation} \label{eq:4}
\rho_s(T) = \frac{\lambda^2 (0)}{\lambda^2 (T)} = \frac{H_{c1}(T)}{H_{c1}(0)}
\end{equation}

The temperature dependence of the $\rho_s$ for Nb$_{0.18}$Re$_{0.82}$ superconductor is shown in Fig. \ref{fig:SFD}. The open symbols with error bars represents the experimental data points. Nb$_{0.18}$Re$_{0.82}$ is a multiband (two-gaps) superconductor as reported in Ref. \cite{cir15}. Hence, we used the two-gap model to explain the superfluid density. For a two-gap superconductor, the normalized superfluid density may be expressed by the following relation \cite{ss15a}. 

\begin{equation} \label{eq:5}
\rho_s (T) = 1 + 2 \left(\int_{\Delta_S (T)}^{\infty} {\frac{dF(E)}{dE} D_S(E)dE} + (1-c)\int_{\Delta_L (T)}^{\infty}{\frac{dF(E)}{dE} D_L(E)dE}\right)
\end{equation}

where $\Delta_S$ and $\Delta_L$ are the small and large superconducting gap, respectively. The parameter $c$ is the fraction that the small gap contributes to the superconductivity. Equation (\ref{eq:5}), is used to fit the normalized superfluid density data, which is shown as the solid line in Fig. \ref{fig:SFD}. The data is well explained using the two-gap model for superfluid density. It is observed that the value of the smaller energy gap is equal to the BCS theoretical limit of 1.76k$_B$$T_c$ and the value of the larger energy gap is greater than the BCS limit. The obtained fitted parameters are shown in Fig. \ref{fig:SFD}. The fitted parameters are well matched with the parameters obtained in Ref. \cite{cir15} from specific heat measurement (dashed line in Fig. \ref{fig:SFD}). Our study provides further confirmation that Nb$_{0.18}$Re$_{0.82}$ is a two-gap superconductor as previously observed in Ref. \cite{cir15}. 

\begin{figure}
\centering
\includegraphics[height=12cm]{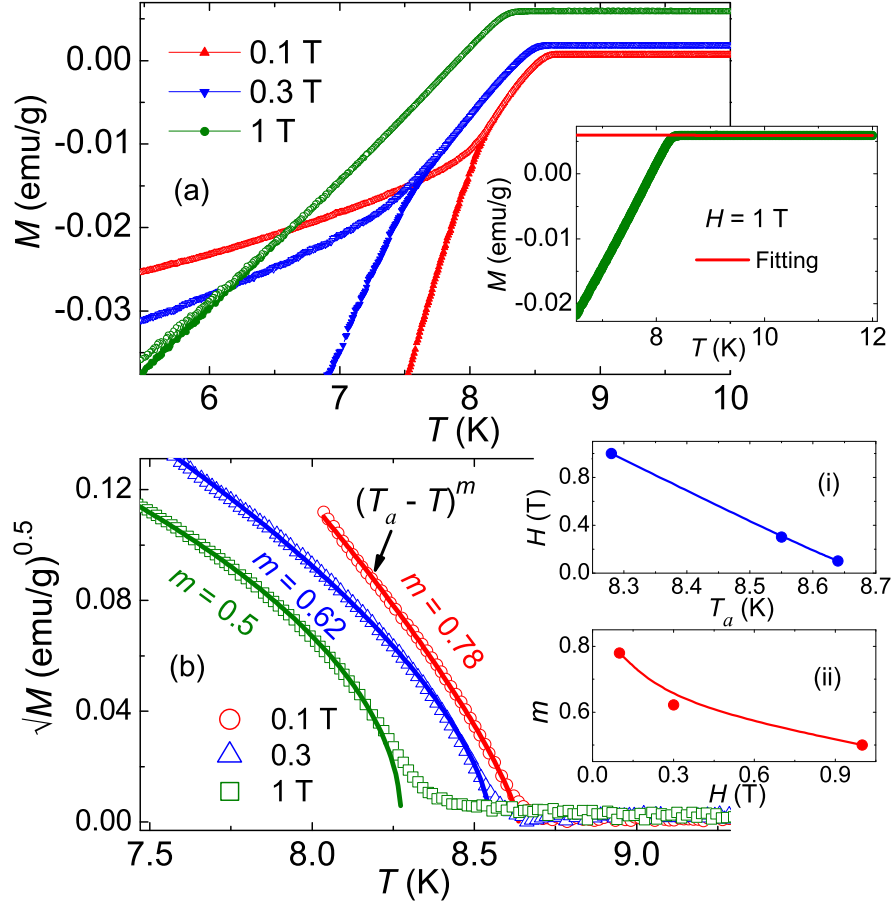}
\caption{\label{fig:MT-rev} (a) Temperature dependence of magnetization in zfc and fcc protocols for different magnetic fields. The inset shows a straight line fit to the experimental data, for $H$ = 1 T, in the normal state, extended down to the lowest temperature to subtract the background contribution in the superconducting state. (b) Isofield curves of $\sqrt{M}$ vs. $T$. Each isofield curve shows a fit to the experimental data using, $\sqrt{M} \propto (T_a(H)-T)^m$. The extracted value of $T_a$ and $m$ are plotted with magnetic field in the inset (i) and (ii) respectively. }
\end{figure}

The search for a spin-triplet component in non-centrosymmetric superconductors is of great interest at present. The non-centrosymmetric lattice structure leads to an antisymmetric spin-orbit coupling (ASOC) in the system. The strength of the ASOC is responsible for the component of spin-triplet pairing, which leads to line nodes in the superconducting energy gap \cite{hay06}. The phase and the amplitude fluctuations of the superconducting order parameter have different contributions in the node and anti-node regions, which may alter the density of states near $T_c$ \cite{hyo99}. Consequently, it may modify the behavior of the superconducting order parameter when compared to the conventional mean field dependence ($\sim$ $(T-T_c)^{1/2}$) near $T_c$. The amplitude and the phase fluctuation analysis relies on the isofield  reversible  magnetization  data, $M(T)$, and is a convenient tool to investigate the non-s-wave behavior in unconventional superconductors \cite{bad10, said08, said09}. 

According to the conventional theory of the upper critical field, $H_{c2}$ \cite{abr57}, the magnetic induction, $B$, obtained from the Ginzburg-Landau (GL) equation may be expressed as \cite{deg89}.

\begin{equation} \label{eq:6}
B = H - \frac{4\pi e\hbar}{mc} {\lvert \psi \rvert}^2
\end{equation}
Where, $\psi$ is the order parameter. Using, $M = \frac{(B-H)}{4\pi}$, we may write

\begin{equation} \label{eq:7}
M = - \frac{e\hbar}{mc} {\lvert \psi \rvert}^2
\end{equation}

This relation shows that the $\sqrt{M}$ is directly proportional to the amplitude of the superconducting order parameter. Hence, near $T_c$, the magnetization may be expressed as, $\sqrt{M} \propto [T_c(H)-T]^m$, where, $T_c(H)$ is the mean field transition temperature. Within the GL theory, for both s-wave and d-wave superconductors, the exponent $m$ is equal to 1/2 \cite{deg89, ji95}. It is known that in low $T_c$ superconductors, the phase of the order parameter is unimportant and the superconducting transition is well described using mean-field theory with an exponent, $m$ = 1/2 \cite{said08}. On the other hand, superconductors with small superfluid density, such as, high $T_c$ oxides, bears a relatively small phase stiffness and poor screening, which leads to  phase fluctuations playing a significant role \cite{eme95}. It is thought that, due to the spin-triplet component, the presence of line nodes may make the phase fluctuations relevant in NCS \cite{bad10}. Consequently, the exponent $m$ may differ significantly from the mean-field value, as observed in the non-centrosymmetric Li$_2$(Pd-Pt)$_3$B superconductor \cite{bad10}.  

Fig. \ref{fig:MT-rev}(a), shows the isofield temperature dependence of the magnetization, $M(T)$, in the zero field cooled (zfc) and field cooled cooling (fcc) protocols, which are used to obtain the reversible (equilibrium) magnetization. Each isofield $M(T)$ curve in Fig. \ref{fig:MT-rev}(a) is corrected for background magnetization using a linear relation, as demonstrated in the inset of Fig. \ref{fig:MT-rev}(a) for $H$ = 1T. Fig. \ref{fig:MT-rev}(b), shows the temperature dependence of $\sqrt{M}$, where, $M$ represents the background corrected value. We are primarily focusing on the reversible region below $T_c(H)$ to analyze the phase fluctuations behavior. The region above $T_c(H)$ is important to study the amplitude fluctuations, which generates an anomalous enhancement of magnetization above $T_c(H)$. The phase-mediated superconducting transition is well described by fitting the reversible magnetization region below $T_c(H)$using the relation, $\sqrt{M} \propto (T_a(H)-T)^m$, where, $T_a(H)$ is the apparent transition temperature and $m$ is the fitting exponent \cite{bad10, said08, said09}. Deviation of $m$ from the mean field value ($m$ = 1/2), suggests the phase-mediated transition. Results for fitting of each isofield $\sqrt{M}$ vs. $T$ curve is shown in Fig. \ref{fig:MT-rev}(b). For $H$ = 0.1 T, the extracted value of the exponent, $m$ = 0.78, which is larger than the mean field value. This suggests that the superconducting transition is phase-mediated, indicating the possibility of a spin-triplet component \cite{bad10}. However, the inset (ii) of Fig. \ref{fig:MT-rev}(b) shows that the exponent, $m$, decreases with applied magnetic field and depicts the mean field value, $m$ = 0.5 for $H$ = 1 T. We speculate that for high magnetic fields and at low temperatures, the lowest Landau levels (LLL) could mask the effect of phase-fluctuations resulting in the mean field type transition being the correct description at high magnetic field. Inset (i) of Fig. \ref{fig:MT-rev}(b), shows the linear variation of apparent transition temperature $T_a$ with magnetic field $H$. The phase-fluctuation analysis suggests the possible admixture of spin-singlet and spin-triplet pairing component in the Nb$_{0.18}$Re$_{0.82}$ non-centrosymmetric superconductor.

\section{Summary and Conclusion}

In conclusion, we presented a detailed investigation of the Nb$_{0.18}$Re$_{0.82}$ superconductor using electrical transport, specific heat and magnetization measurements. Structural characterization is performed using x-ray diffraction and optical metallography techniques, both confirm the high quality of the sample. Electrical resistivity in the normal state is interpreted using an empirical relation, which includes the presence of phonon-assisted inter-band scattering. The magnetoresistance does not follow the conventional Kohler's rule and supports the two-band model. The upper critical field, $H_{c2}$, does not follow the Werthamer, Helfand and Hohenberg (WHH) model in the whole temperature range of the measurements and the zero temperature limit of $H_{c2}$ exceeds the Pauli limit, both suggesting the possibility of unconventional pairing. The Kadowaki-Woods and the Uemura plots do not support a conventional behavior as claimed previously. The estimated lower critical field, $H_{c1}$, follows an unexpected cubic temperature dependence and the derived normalized superfluid density is well explained with the two-gap picture. The phase of the superconducting order parameter is analysed using the reversible magnetization data, which indicates that the superconducting transition is phase mediated. This implies the possibility of an admixture of spin singlet and spin triplet pairing consistent with the anticipated influence of antisymmetric spin-orbit coupling in this non-centrosymmetric superconductor.  

\section*{Acknowledgements}

Authors would like to thank Mr. R. K. Meena for assistance in sample preparation and Alberto A. Mendon{\c c}a for assistance in XRD measurements. SS acknowledges a post-doctoral fellowship from FAPERJ (Rio de Janeiro, Brazil), processo: E-26/202.848/2016. SSS and LG are supported by CNPq and FAPERJ. 

\section*{}

$^*$ Corresponding author: shyam.phy@gmail.com \newline
$^\ddagger$ Currently superannuated


\bibliographystyle{unsrt}



\end{document}